\documentstyle[prl,aps,multicol]{revtex} 
 
\draft 
 
\author{M.~A.~Nielsen$^{1,}$\thanks{Email: nielsen@physics.uq.edu.au}
 and J.~Kempe$^{2,}$\thanks{also \'Ecole Nationale Superieure des
    T\'el\'ecommunications, Paris, France, Email: kempe@math.berkeley.edu}}
\title{Separable states are more disordered globally than locally} 
\address{$^1$Center for Quantum Computer Technology, 
  University of Queensland 4072, Australia\\
$^2$ Departments of Mathematics and Chemistry, University of California, Berkeley 
} 
\date{\today} 
 
\begin{document}

\pagestyle{plain} 
\pagenumbering{arabic} 
 
\maketitle 
 
\begin{abstract} 
  A remarkable feature of quantum entanglement is that an entangled
  state of two parties, Alice ($A$) and Bob ($B$), may be more
  disordered locally than globally.  That is, $S(A) > S(A,B)$, where
  $S(\cdot)$ is the von~Neumann entropy.  It is known that
  satisfaction of this inequality implies that a state is
  non-separable.  In this paper we prove the stronger result that for
  separable states the vector of eigenvalues of the density matrix of
  system $AB$ is majorized by the vector of eigenvalues of the density
  matrix of system $A$ alone.  This gives a strong sense in which a
  separable state is more disordered globally than locally and a new
  {\em necessary} condition for separability of bipartite states in
  arbitrary dimensions. We also investigate the extent to which these
  conditions are {\em sufficient} to characterize separability,
  exhibiting examples that show separability cannot be characterized
  solely in terms of the local and global spectra of a state.  We
  apply our conditions to give a simple proof that non-separable
  states exist sufficiently close to the completely mixed state of $n$
  qudits.
\end{abstract} 
 
 
\begin{multicols}{2}[] 
\narrowtext 
 
%
%
Quantum mechanics harbours a rich structure whose investigation and
explication is the goal of quantum information
science\cite{Nielsen00a,Preskill98c}.  At present only a limited
understanding of the fundamental static and dynamic properties of
quantum information has been obtained, and many major problems remain
open.  In particular, we would like a detailed ontology and
quantitative methods of description for the different types of
information and dynamical processes afforded by quantum mechanics.  An
example of the pursuit of these goals has been the partial development
of a theory of quantum entanglement; see,
e.g.,~\cite{Bennett96a,Nielsen99a,Horodecki96f,Vedral98a,Vidal99a} and
references therein.
 
%
%
The {\em separability} or non-separability of a quantum state is a
question that has received much attention in the development of a
theory of entanglement.  The notion of separability captures the idea
that a quantum state's static properties can be explained entirely by
classical statistics, and is sometimes claimed to be equivalent to the
notion that a state is ``not entangled''.  More precisely, a state
$\rho_{AB}$ of Alice and Bob's system is separable\cite{Werner89a} if
it can be written in the form $\rho_{AB} = \sum_j q_j \rho_j \otimes
\sigma_j$, for some probability distribution $\{q_j\}$, and density
matrices $\rho_j$ and $\sigma_j$ of Alice and Bob's systems,
respectively.  Thus, we can think of Alice and Bob's systems as having
a local, pseudo-classical description, as a mixture of the product
states $\rho_j \otimes \sigma_j$ with probabilities $q_j$.  Note that
separability is equivalent to the condition
\begin{eqnarray} \label{eq:sep}
  \rho_{AB} = \sum_j p_j |\psi_j\rangle \langle \psi_j| \otimes 
  |\phi_j \rangle \langle \phi_j|, 
\end{eqnarray} 
where $\{ p_j \}$ is a probability distribution and $|\psi_j\rangle,
|\phi_j\rangle$ are pure states of Alice and Bob's systems,
respectively.
 
%
%
One reason for interest in separability is a deep theorem due to M.,
P.~and R.~Horodecki connecting separability to positive maps on
operators\cite{Horodecki96f}.  The Horodeckis used this theorem to
prove that the ``positive partial transpose'' criterion for
separability introduced by Peres\cite{Peres96b} is a necessary and
sufficient condition for separability of a state $\rho_{AB}$ of a
system consisting of a qubit in Alice's possession, and either a qubit
or qutrit in Bob's possession.  More precisely, if we define
$\rho^{T_B}_{AB}$ to be the operator that results when the
transposition map is applied to system $B$ alone, then the Horodeckis
showed that $\rho_{AB}$ is separable if and only if $\rho^{T_B}_{AB}$
is a positive operator.  Unfortunately, this criterion, while
necessary for a state to be separable in higher
dimensions\cite{Peres96b}, is not sufficient.
 
%
%
A hallmark of quantum entanglement is the remarkable fact that
individual components of an entangled system may exhibit {\em more}
disorder than the system as a whole.  The canonical example of this
phenomenon is a pair of qubits $A$ and $B$ prepared in the maximally
entangled state $(|00\rangle+|11\rangle)/\sqrt 2$.  The von~Neumann
entropy $S(A)$ of qubit $A$ is equal to one bit, compared with a
von~Neumann entropy $S(A,B)$ of zero bits for the joint system.
Classically, of course, such behaviour is impossible, and the Shannon
entropy $H(X)$ of a single random variable is never larger than the
Shannon entropy of two random variables, $H(X),H(Y) \leq H(X,Y)$.  It
has been shown~\cite{Horodecki96c} (see Chapter~8 of~\cite{Nielsen98d}
and~\cite{Cerf99a,Horodecki98c} for related results) that an analogous
relation holds for separable states,
\begin{eqnarray} 
  \label{eq:entropy_cond} 
  S(A), S(B) \leq S(A,B).
\end{eqnarray} 
This result is a consequence of the concavity of
$S(A,B)-S(A)$~\cite{Lieb73b,Nielsen00a}, since when $\rho_{AB} =
\sum_j q_j \rho_j \otimes \sigma_j$ we have $S(A,B)-S(A) \geq \sum_j
q_j (S(\rho_j\otimes \sigma_j)-S(\rho_j)) \geq 0$.  Unfortunately, the
inequalities~(\ref{eq:entropy_cond}) are insufficient to characterize
separability.  To see this, consider the Werner state of two qubits
$\rho_p = p|\Psi\rangle \langle \Psi|+(1-p)I/4$ ($0 \leq p \leq 1$)
and $|\Psi\rangle = (|00\rangle+|11\rangle)/\sqrt 2$.  The positive
partial transpose criterion implies that the state is separable iff $p
\leq 1/3$.  The marginal density matrices being fully mixed for all
$p$, however, one obtains $S(A)=S(B)=1 \leq
S(A,B)=H(\frac{1+3p}{4},\frac{1-p}{4},\frac{1-p}{4},\frac{1-p}{4})$
for $0 \leq p \leq 0.747...$, so the condition~(\ref{eq:entropy_cond})
is fulfilled for a range of inseparable states.
 
The notion of von~Neumann entropy is a valuable notion of disorder in
a quantum state, however more sophisticated tools for quantifying
disorder exist.  One such tool is the theory of majorization, whose
basic elements we now review (see Chapters~2 and~3
of~\cite{Bhatia97a}, \cite{Marshall79a} or~\cite{Alberti82a} for more
extensive background).  Suppose $x = (x_1,\ldots,x_d)$ and
$y=(y_1,\ldots,y_d)$ are two $d$-dimensional real vectors; we usually
suppose in addition that $x$ and $y$ are probability distributions,
that is, the components are non-negative and sum to one.  The relation
$x \prec y$, read ``$x$ is majorized by $y$'', is intended to capture
the notion that $x$ is more ``mixed'' (i.e.  disordered) than $y$.
Introduce the notation $\downarrow$ to denote the components of a
vector rearranged into decreasing order, so $x^{\downarrow} =
(x_1^{\downarrow},\ldots,x_d^{\downarrow})$, where $x_1^{\downarrow}
\geq x_2^{\downarrow} \geq \ldots \geq x_d^{\downarrow}$.  Then we
define $x \prec y$, if
\begin{eqnarray} 
  \sum_{j=1}^k x_j^{\downarrow} \leq \sum_{j=1}^k y_j^{\downarrow}, 
\end{eqnarray} 
for $k=1,\ldots,d-1$, and with the inequality holding with equality
when $k = d$.  To understand how this definition connects with
disorder consider the following result (see Chapter~2 of
\cite{Bhatia97a} for a proof): $x \prec y$ if and only $x = Dy$, where
$D$ is a doubly stochastic matrix.  Thus, when $x \prec y$ we can
imagine that $y$ is the input probability distribution to a noisy
channel described by the doubly stochastic matrix $D$, inducing a more
disordered output probability distribution, $x$.  Majorization can
also be shown~\cite{Bhatia97a} to be a more stringent notion of
disorder than entropy in the sense that if $x \prec y$ then it follows
that $H(x) \geq H(y)$.

%
%
Given the known connections between measures of disorder such as the
von~Neumann entropy and separability, it is natural to conjecture that
there might be some relationship between separability and the vectors
$\lambda(\rho_{AB}), \lambda(\rho_{A}), \lambda(\rho_B)$ of
eigenvalues for $\rho_{AB}$ and the corresponding reduced density
matrices.  Majorization suggests the following theorem as a natural
way of strengthening the necessary conditions for
separability, Eqtn.~(\ref{eq:entropy_cond}):

{\em Theorem 1:}
  \label{eq:theorem} 
{\em If $\rho_{AB}$ is separable then} 
\begin{equation} \label{eq:crit}
\lambda(\rho_{AB}) \prec 
  \lambda(\rho_A) \quad \mbox{{\em and}} \quad \lambda(\rho_{AB}) \prec \lambda(\rho_B).
\end{equation}
(By convention we append zeroes to the vectors $\lambda(\rho_A)$ and
$\lambda(\rho_B)$ so they have the same dimension as
$\lambda(\rho_{AB})$.)

%
%
Theorem~1 is the main result of this paper.  Note that it provides a
more stringent criterion for separability
than~(\ref{eq:entropy_cond}), since for any two states $\rho$ and
$\sigma$, $\lambda(\rho) \prec \lambda(\sigma)$ implies that
$S(\rho) \geq S(\sigma)$, but not necessarily conversely.

%
%
{\em Proof:} If $\rho_{AB}$ is separable, it may be written in the
form of~(\ref{eq:sep}).  Let $\rho_{AB} = \sum_k r_k |e_k\rangle
\langle e_k|$ be a spectral decomposition for $\rho_{AB}$.  By the
classification theorem for ensembles (Theorem~2.6 in
\cite{Nielsen00a}) it follows that there is a unitary matrix $u_{kj}$
such that
\begin{eqnarray} 
  \label{eq:thm_inter} 
  \sqrt{r_k} |e_k\rangle & = & \sum_j u_{kj} \sqrt{p_j} 
  |\psi_j\rangle |\phi_j\rangle. 
\end{eqnarray} 
Next we trace out system $B$ in~(\ref{eq:sep}) to give $\rho_A =
\sum_j p_j |\psi_j\rangle \langle \psi_j|$.  Letting $\rho_A = \sum_l
a_l |f_l\rangle \langle f_l|$ be a spectral decomposition and applying
the classification theorem for ensembles we see that there is a
unitary matrix $v_{jl}$ such that $\sqrt{p_j} |\psi_j\rangle = \sum_l
v_{jl} \sqrt{a_l} |f_l\rangle$.  Substituting
into~(\ref{eq:thm_inter}) gives $\sqrt{r_k} |e_k\rangle = \sum_{jl}
\sqrt{a_l} u_{kj} v_{jl} |f_l\rangle |\phi_j\rangle$.  Multiplying
this equation by its adjoint and using the orthonormality of the
vectors $|f_l\rangle$ we obtain
\begin{eqnarray} 
  r_k = \sum_l D_{kl} a_l. 
\end{eqnarray}
where  
\begin{eqnarray} 
  \label{eq:D_defn} 
  D_{kl} \equiv \sum_{j_1 j_2} u_{kj_1}^* u_{kj_2} v_{j_1l}^* v_{j_2l} 
  \langle \phi_{j_1}|\phi_{j_2}\rangle.
\end{eqnarray}  
To complete the proof all we need to do is show that $D_{kl}$ is
doubly stochastic.  The fact that $D_{kl} \geq 0$ follows by defining
$|\gamma_{kl} \rangle \equiv \sum_j u_{kj}v_{jl} |\phi_j\rangle$ and
noting that $D_{kl} = \langle \gamma_{kl} |\gamma_{kl}\rangle \geq 0$.
From~(\ref{eq:D_defn}) and by the unitarity of $u$ we have
\[  \sum_k D_{kl} =  \sum_{j_1 j_2} \delta_{j_1 j_2} v_{j_1 l}^* 
  v_{j_2l} \langle \phi_{j_1}|\phi_{j_2}\rangle  =  \sum_j v_{jl}^* v_{jl} =1.\] 
Similarly, $\sum_l D_{kl} = 1$, and thus $D$ is a doubly stochastic 
matrix. $\Box$

%
%
The separability criterion~(\ref{eq:crit}) is strictly stronger than
the entropic criterion~(\ref{eq:entropy_cond}).  Indeed, for
Bell-diagonal states of two qubits, it follows from the positive
partial transpose criterion and a straightforward calculation that
condition~(\ref{eq:crit}) is equivalent to separability, whereas as
remarked earlier the condition $S(A), S(B) \leq S(A,B)$ is not
sufficient to characterize separability even for the more restricted
case of Werner states.  More generally, the separability
criterion~(\ref{eq:crit}) completely characterizes the separability
properties of Werner states in arbitrary ($d$) dimensions.  More
precisely, states of the form $\rho_p=p |\Psi\rangle \langle \Psi| +
(1-p)/d^2 I$ where $|\Psi\rangle=(|00\rangle+|11\rangle + \ldots +
|(d-1)(d-1)\rangle)/\sqrt{d}$ are known to be separable iff $p \leq
1/(d+1)$ \cite{Dur00c}.  The marginal density matrices of these states
are completely mixed and the criterion~(\ref{eq:crit}) thus becomes
\begin{eqnarray}
\frac{1}{d^2}(1+(d^2-1)p,1-p,\ldots,1-p)  \prec
\frac{1}{d}(1,\ldots,1),
\end{eqnarray}
which is easily seen to be equivalent to $p\leq 1/(d+1)$.

%
%
Another interesting application of the conditions~(\ref{eq:crit}) is
to the problem of finding non-separable states near the completely
mixed state $I^{\otimes n}/d^n$ of $n$ qudits ($d$-dimensional quantum
systems).  Consider the state $\rho \equiv (1-\epsilon)I^{\otimes
  n}/d^n +\epsilon|\psi\rangle \langle \psi|$, where $|\psi\rangle$ is
the cat state of $n$ qudits.  Partitioning the $n$ qudits so that the
first $n-1$ belong to Alice, and the final qudit to Bob, a
straightforward calculation shows that the conditions~(\ref{eq:crit})
are violated whenever $\epsilon > 1/(1+d^{n-1})$, and thus $\rho$ must
be inseparable when $\epsilon$ satisfies this condition.  Note that
this result has previously been obtained by other
techniques~\cite{Pittenger00a,Rungta00a} (see
also~\cite{Dur99a,Vidal99b,Caves00a,Braunstein99a}), however the
utility of the conditions~(\ref{eq:crit}) is demonstrated in this
application by the ease with which they are applied and their
generality, as compared to the more complex and state-specific
arguments used previously to study the separability of $\rho$.

%
%
It is natural to conjecture the converse to Theorem~1, that if both
the conditions in~(\ref{eq:crit}) hold then $\rho_{AB}$ is separable.
Unfortunately, this is not the case, as the following two
qubit example shows.  \\
{\em Example 1:} Let $\rho_{AB}^p \equiv p |00\rangle \langle 00| +
(1-p)|\Phi\rangle \langle \Phi|$ with the Bell state
$|\Phi\rangle=(|01\rangle+|10\rangle)/\sqrt 2$.  Then the partial
transpose criterion implies that this state is non-separable whenever
$p \neq 1$. However $\lambda(\rho_{AB}^p) = (p,1-p) \prec
\lambda(\rho^p_{A,B}) = ((1+p)/2,(1-p)/2)$ for $1/3 \leq p$, that is,
criterion~(\ref{eq:crit}) is fulfilled for this non-separable state.

%
%
More generally, we now show that attempts to characterize separability
based only upon the eigenvalue spectra $\lambda(\rho_{AB}),
\lambda(\rho_A)$ and $\lambda(\rho_B)$ can never work.  We will
demonstrate this by exhibiting a pair of two qubit states $\rho_{AB}$
and $\sigma_{AB}$ such that all these vectors of eigenvalues are the
same (i.e., the states are globally and locally {\em isospectral}),
yet $\rho_{AB}$ is not separable, while $\sigma_{AB}$ is.  \\
{\em Isospectral Example:}
\begin{equation} \label{eq:iso}
  \rho_{AB}  =  \frac{1}{3} \left[ \begin{array}{cccc}  
      1 & 0 & 0 & 0 \\ 
      0 & 1 & 1 & 0 \\ 
      0 & 1 & 1 & 0 \\ 
      0 & 0 & 0 & 0 
      \end{array} \right] \quad 
  \sigma_{AB}  =  \left[ \begin{array}{cccc} 
      \frac{1}{3} & 0 & 0 &           0 \\ 
                0 & 0 & 0 &           0 \\ 
                0 & 0 & 0 &           0 \\ 
                0 & 0 & 0 & \frac{2}{3} 
      \end{array} \right] 
\end{equation} 
The isospectrality of the these states may be checked by direct
calculation, and the fact that $\rho_{AB}$ is non-separable while
$\sigma_{AB}$ is follows from the partial transpose criterion.  (Note
that similar examples have also been found by Richard~Davis (private
communication).)  It is worth emphasizing how remarkable such examples
are: these density matrices have the same spectra, both globally and
locally, yet one is separable, while the other is not.  This runs
counter to the often-encountered wisdom that a complete understanding
of a quantum system can be obtained by studying the local and global
properties of the spectra of that system.  This is the point of view
apparently adopted, for instance, in the theory of quantum phase
transitions\cite{Sachdev99a}, perhaps leading to the disregard of
important physical effects in that theory.

%
%
Given the isospectral example it is natural to ask under what
conditions a separable state exists, given specified global 
and local spectra.  We can report the following result in this direction.\\
{\em Theorem 2:} If $\rho_{AB}$ is a density matrix such that
$\lambda(\rho_{AB}) \prec \lambda(\rho_A)$, then there exists a
separable density matrix $\sigma_{AB}$ such that $\lambda(\sigma_{AB})
= \lambda(\rho_{AB})$ and $\lambda(\sigma_{A}) =
\lambda(\rho_{A})$.\\
{\em Proof:} Suppose $(r_j) = \lambda(\rho_{AB})$ and $(s_k) =
\lambda(\rho_A)$.  By Horn's lemma\cite{Horn54a,Nielsen00c}, there is
a unitary matrix $u_{jk}$ such that $s_j = \sum_k |u_{jk}|^2 r_k$.
Introduce orthonormal bases $|j\rangle$ for system $B$ and $|k\rangle$
for system $A$, and for each non-zero $r_j$ define
\begin{eqnarray} 
  |\psi_j\rangle \equiv \frac{\sum_k u_{jk} \sqrt{s_k} |k\rangle}{\sqrt{r_j}}. 
\end{eqnarray} 
Then define $\sigma \equiv \sum_j r_j |\psi_j \rangle \langle \psi_j|
\otimes |j\rangle \langle j|$.  Note that $\sigma$ is manifestly
separable with spectrum $\lambda(\rho_{AB})$, while a simple
calculation shows that $\mbox{tr}_B(\sigma) = \sum_k s_k |k\rangle
\langle k|$, and thus $\lambda(\sigma_A) = \lambda(\rho_A)$,
completing the proof. $\Box$

%
%
A stronger conjecture is that whenever {\em both} $\lambda(\rho_{AB})
\prec \lambda(\rho_A)$ {\em and} $\lambda(\rho_B)$, then there exists
a separable state $\sigma_{AB}$ which is isospectral to $\rho_{AB}$.
Unfortunately, the following theorem shows
that this is not true.\\
{\em Theorem 3:} For the class of states $\rho_{AB}^p$ in Example~1
(which are non-separable when $1> p > 1/3$) the separability
conditions~(\ref{eq:crit}) are fulfilled yet there is no separable
$\sigma_{AB}$ (globally and locally) isospectral to $\rho_{AB}^p$ when
$1 > p \geq 1/2$. \\
%
%
{\em Proof:} Suppose $\sigma \equiv \sigma_{AB}$ is a separable state
isospectral to $\rho_{AB}^p$.  Then $\sigma = p|s_1\rangle \langle
s_1|+(1-p)|s_2\rangle \langle s_2|$ for orthonormal states
$|s_1\rangle$ and $|s_2\rangle$.  We suppose for now that $\sigma$ can
be given a separable decomposition with only two terms, $\sigma =
q|a_1\rangle \langle a_1| \otimes |b_1\rangle \langle b_1| + (1-q)
|a_2\rangle \langle a_2| \otimes |b_2\rangle \langle b_2|$.  We show
later that this is the only case that need be considered.  Define
angles $\alpha, \beta$ and $\phi$ by $|\langle a_1| b_1 \rangle |
\equiv \cos(\alpha); |\langle a_2|b_2\rangle | \equiv \cos(\beta);
\cos(\phi) \equiv \cos(\alpha)\cos(\beta)$.  Then the global and local
spectra for $\sigma$ are easily calculated,
\begin{eqnarray}
  \lambda(\sigma_{AB}) = \left( \frac{1 \pm
  \sqrt{1-4q(1-q)\sin^2(\phi)}}{2} \right),
\end{eqnarray}
with similar expressions for $\lambda(\sigma_A)$ and
$\lambda(\sigma_B)$, with $\alpha$ and $\beta$ appearing in place of
$\phi$.  Assuming $1/2 \leq p$ this gives $\sin^2(\alpha) =
\sin^2(\beta) = (1-p^2)/4q(1-q)$ and $p(1-p) = q(1-q)\sin^2(\phi)$.
Using $\sin^2(\phi) = 1-(1-\sin^2(\alpha))(1-\sin^2(\beta))$ to
substitute the former expression into the latter, we find $q(1-q) =
(1+p)^2/8$.  For $p > \sqrt{2}-1 \approx 0.41$ there is no $q$ in the
range $0$ to $1$ satisfying this equation, so we deduce that no such
separable state $\sigma$ can exist.

%
%
To complete the proof we show that any separable decomposition $\sigma
= \sum_j q_j |a_j\rangle \langle a_j| \otimes |b_j\rangle \langle
b_j|$ can be assumed to have two terms.  Without loss of generality we
assume that there is no redundancy in the decomposition, that is,
there do not exist values $j \neq k$ such that $|a_j\rangle|b_j\rangle
= |a_k\rangle|b_k\rangle$ (up to phase).  We show that assuming the
decomposition has three or more terms leads to a contradiction.  Note
that the decomposition must contain contributions from at least two
linearly independent states, say $|a_1\rangle|b_1\rangle$ and
$|a_2\rangle |b_2\rangle$.  Furthermore, because $\mbox{rank}(\sigma)
= 2$ any other state in the sum must be a linear combination of these
two states, $|a_j\rangle |b_j\rangle = \alpha_j|a_1\rangle |b_1\rangle
+ \beta_j|a_2\rangle |b_2\rangle$.  By the no-redundancy assumption
neither $|\alpha_j| = 1$ nor $|\beta_j| = 1$, so we must have $0 <
|\alpha_j|, |\beta_j| < 1$.  Consider now three possible cases.  In
the first case, $|a_1\rangle = |a_2\rangle$ (up to phase), in which
case $|a_j\rangle = |a_1\rangle$ (up to phase) for all $j$, and thus
$\lambda(\sigma_A) = (1,0) \neq \lambda(\rho_A^p)$, a contradiction.
A similar contradiction arises when $|b_1\rangle = |b_2\rangle$ up to
phase.  The third and final case is when neither $|a_1\rangle =
|a_2\rangle$ nor $|b_1\rangle = |b_2\rangle$ up to phase.  In this
case $\alpha_j|a_1\rangle|b_1\rangle + \beta_j |a_2\rangle
|b_2\rangle$ cannot be a product state, a contradiction. $\Box$

%
%
Given that attempts to characterize separability based on the local
and global spectra are doomed to failure, it is still interesting to
ask whether the conditions $\lambda(\rho_{AB}) \prec \lambda(\rho_A)$
and $\lambda_(\rho_{AB}) \prec \lambda(\rho_B)$ are equivalent to some
other interesting physical condition?  We have tried to find such an
equivalence, with little success, but can identify several plausible
possibilities which these conditions are {\em not} equivalent to.
They are not equivalent to the property of violating a Bell
inequality, of having a positive partial transpose, or of being
distillable.
%
%
Another interesting idea is to find states which have positive partial
transposition, but which violate~(\ref{eq:crit}).  Such a state will
necessarily be bound-entangled \cite{Horodecki98b}.  We have not yet
identified any such states, despite searching through several of the
known classes of bound-entangled states, and doing numerical searches.

%
%
In summary, we have connected two central notions in the theory of
entanglement, using majorization to obtain a simple set of necessary
conditions for a state to be separable in arbitrary dimensions.
Understanding the physical import of these conditions and their
relationship to criteria such as the positive partial transpose
conditon remains an interesting problem for further research.
 
\section*{acknowledgments} 
Thanks to Carl~Caves and Pranaw Rungta for discussions about
separability, and quantum information in general. JK thanks the Center
for Quantum Computer Technology at the University of Queensland for
its hospitality, and acknowledges support by the U.S.~ARO under
contract/grant number DAAG55-98-1-0371.

\end{multicols} 
 
\end{document}